\documentclass{iauc} 
\usepackage{graphicx}
\usepackage{multicol}
\pubyear{2005} \volume{199} \pagerange{1--} 
\jname{Probing Galaxies through Quasar Absorption Lines}
\editors{P. R. Williams, C. Shu, and B. M\'{e}nard, eds.}

\title[Nature of Mg\,II Absorbers]    
{Towards an Understanding of the Physical Nature of Mg\,II Absorption Systems}

\author[Nestor, Turnshek \& Rao]   
{D. B. Nestor$^1$, D. A. Turnshek$^2$ \and S. M. Rao$^2$}

\affiliation{$^1$Department of Astronomy, University of Florida,
Gainesville, FL 32611, USA \break  email:
dbn@astro.ufl.edu\\[\affilskip] $^2$Department of Physics \&
Astronomy, University of Pittsburgh, \break Pittsburgh, PA 15260, USA
\break  email: turnshek@quasar.astro.pitt.edu,
rao@everest.phyast.pitt.edu}

\begin{document}

\maketitle

\begin{abstract}
We discuss issues concerning the physical nature of intervening Mg\,II
quasar absorption systems in light of results from our recent surveys
using SDSS EDR QSO spectra and data obtained at the MMT.  These
surveys indicate an excess of weak ($W_0^{\lambda2796} \lesssim
0.3$\AA) systems relative to the exponential $\partial N/\partial W_0$
distribution of stronger systems.  The incidence of
intermediate-strength lines shows remarkably little evolution with
redshift, thereby constraining models for the nature of the clouds
comprising these absorbers.  The {\it total} distribution does evolve,
with the incidence decreasing with decreasing
redshift in a $W_0^{\lambda2796}$-dependent rate (the strongest
systems evolve the fastest).  This suggests that multiple populations
that evolve at different rates contribute to the incidence in a
$W_0^{\lambda2796}$-dependent manner.  We also present two images of
fields containing unprecedented ``ultra-strong'' ($W_0^{\lambda2796}
\ge 4.0$\AA) Mg\,II absorbers.

\end{abstract}

\firstsection 
\section{Introduction}

\cite{BS77} demonstrated that some quasar absorption lines 
arise in intervening galaxies with the discovery of the Ca\,II H 
and K absorption lines from NGC 3067 in the spectrum of 3C 232.  In the
subsequent decades, much progress has been made in the understanding
of the physical conditions, such as abundances (\cite{N03};
\cite{P03}), temperatures (e.g., \cite{LBS00}) and kinematics
(\cite{CV01}), of the gas responsible for intervening
low-ion/neutral-gas-rich absorption.  However, an overall physical
description of the structures giving rise to the absorption has been
more elusive.  Do these absorbers select extended galaxy disks, halos,
stripped gas, galactic winds, clouds condensing from ionized gas
surrounding galaxies, enriched IGM gas not directly associated with a
specific galaxy, some combination of processes, or do they have other natures
altogether?  More importantly, what does the answer to this question
tell us about the {\it physical nature of galaxies}, and how the properties 
of galaxies {\it evolve with cosmic time}?  In this contribution we
address these issues in light of the results of our recent surveys for
Mg\,II absorbers in Sloan Digital Sky Survey (SDSS) Early Data
Release (EDR) quasar spectra and in data obtained at the 6.5 m 
MMT on Mount Hopkins, AZ.

\section{Incidence of Intervening Mg\,II Absorption Systems}
Perhaps the most fundamental result from absorption line surveys is
the determination of the incidence of absorbers.  As regards Mg\,II
systems, an important early survey was 
\cite{LTW87}. The survey of \cite{SS92} determined
the incidence of systems over the interval $0.2 < z < 2.1$ for
absorbers with Mg\,II $\lambda2796$ rest equivalent width $W_0^{\lambda2796}
\ge 0.3$\AA. \cite{CRCV99}  determined the incidence of ``weak''
($W_0^{\lambda2796} < 0.3$\AA)  systems.

It is both traditional and useful to consider the incidence (proportional
to total cross section, which is the integrated product of absorber
cross section and comoving number density) 
of systems, $\partial^2 N/\partial z \partial W_0$, as it depends on 
$W_0^{\lambda2796}$ and redshift separately.

\subsection{Incidence as a Function of Line Strength}
\begin{figure}
\centerline{ \includegraphics[scale=0.95]{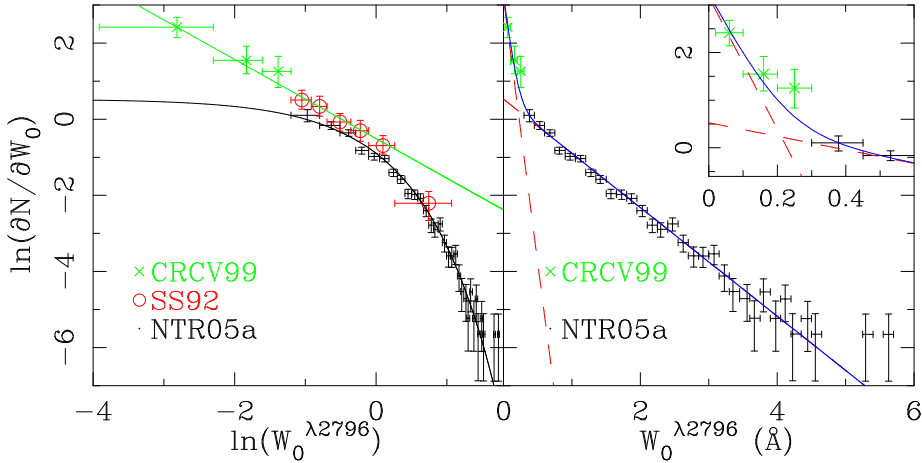}}
  \caption{Dependence of the incidence of Mg\,II absorbers on
$W_0^{\lambda2796}$.  Left: the surveys of SS92, CRCV99 and NTR05a.
Right:  fitting the data with a sum of two exponentials.}
\label{fig:dndw}
\end{figure}

The incidence as a function of Mg\,II $\lambda2796$ line strength,
$\partial N/\partial W_0^{\lambda2796}$, from the SS92 and CRCV99
surveys are shown in the left panel of Figure~\ref{fig:dndw}.
SS92 fitted their data equally well with a power law or an
exponential, but the combined SS92 and CRCV99 results suggested that
$\partial N/\partial W_0^{\lambda2796}$ was best described by a power
law.  In \cite{NTR05a}, we presented the results of a detailed Mg\,II
survey using the SDSS EDR quasar spectra.  Though the redshift
coverage and minimum $W_0^{\lambda2796}$ of this survey were similar
to SS92, it was more than an order of magnitude larger which not only
provided smaller statistical errors but allowed for the measurement of
systems with $W_0^{\lambda2796}$ up to almost 6\AA.  The incidence
determined from this survey, also shown in Figure~\ref{fig:dndw},
indicates that the distribution for $W_0^{\lambda2796} \gtrsim
0.3$\AA\ is very well fit by an exponential.  Power law fits
drastically over predict the incidence of the strongest systems.
However, extrapolating our exponential fit to the weak regime
drastically under predicts the CRCV99 results.

\cite{RCC02} demonstrated that the number of clouds (kinematic
sub-components) comprising a Mg\,II absorber is consistent with a
Poissonian distribution, {\it except} for the excess of single-cloud
systems, which are optically-thin in neutral hydrogen and comprise
$\approx 2/3$ of ``weak'' systems.  It thus appears that these
single-cloud systems are indeed a distinct population.  Therefore, we
fitted the combined $W_0^{\lambda2796}$ incidence data with a sum of
two exponentials, shown in the right panel of Figure~\ref{fig:dndw}.
Indeed, the extrapolation of the shallow exponential accounts for
$\approx 1/3$ of the expected number of weak systems consistent with
the fraction of multi-cloud weak systems found by RCC02.

\begin{figure}
\centerline{ \includegraphics[scale=0.6]{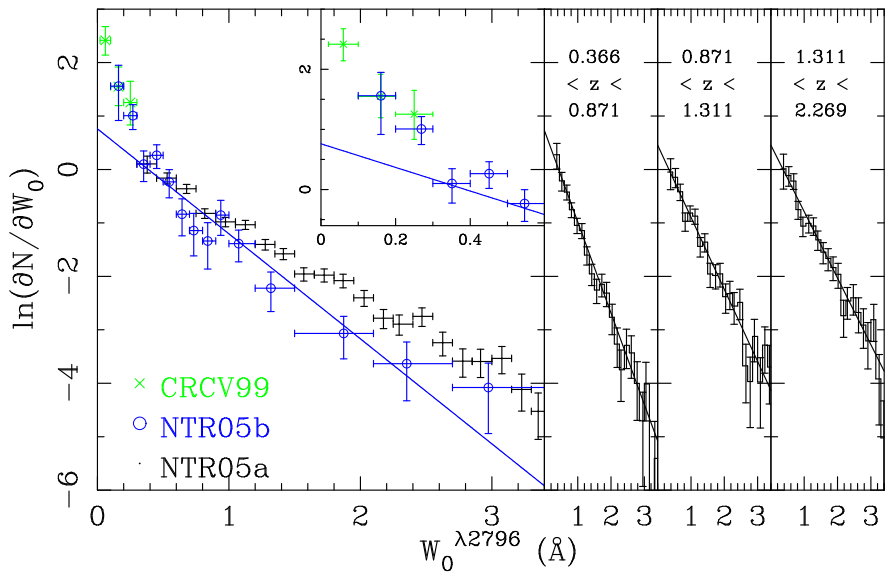}}
  \caption{Left: $\partial N/\partial W_0^{\lambda2796}$ from the EDR,
  MMT and CRCV99 surveys.  Right:  redshift dependence of $\partial
  N/\partial W_0^{\lambda2796}$ from the EDR survey.}
\label{fig:dndw2}
\end{figure}

Since the CRCV99 data does not overlap with the SS92 or NTR05a data,
it is a concern that the upturn below $W_0^{\lambda2796} \approx
0.3$\AA\ is due to some unknown systematic difference in the surveys.
We were able address this concern with our MMT survey, which achieved
greater sensitivity allowing the study of weaker lines.  The incidence
as a function of $W_0^{\lambda2796}$ determined from the MMT survey
\cite{NTR05b} is shown in the left panel of Figure~\ref{fig:dndw2}.
Also shown is the EDR and CRCV99 results.  One can see that the
$W_0^{\lambda2796} < 0.3$\AA\ MMT results are quite consistent with
CRCV99 and significantly above  the extrapolation of the exponential
fit to the stronger systems.  However, the slope of the exponential
measured in the MMT data is steeper
than that determined from the EDR data due to the difference in mean
redshift of the surveys: $\left< z \right>_{MMT} = 0.589$ and $\left<
z \right>_{EDR} = 1.108$.  As can be seen in the right panels of
Figure~\ref{fig:dndw2} which show the EDR results divided into three
redshift ranges, $\partial N/\partial W_0^{\lambda2796}$ steepens with
decreasing redshift.  The MMT results for $W_0^{\lambda2796} >
0.3$\AA\ are consistent with the lower-redshift EDR results.

\subsection{Incidence as a Function of Redshift}
Multi-cloud systems span a large range of strengths, and thus velocity
spreads (since $W_0^{\lambda2796}$ for saturated lines is primarily a
measure of line-of-sight velocity complexity).  Insight into the
nature of the stronger systems can be gained by considering how their
incidence depends on redshift, and thus cosmic time.

\begin{figure}
\centerline{ \includegraphics[scale=0.59]{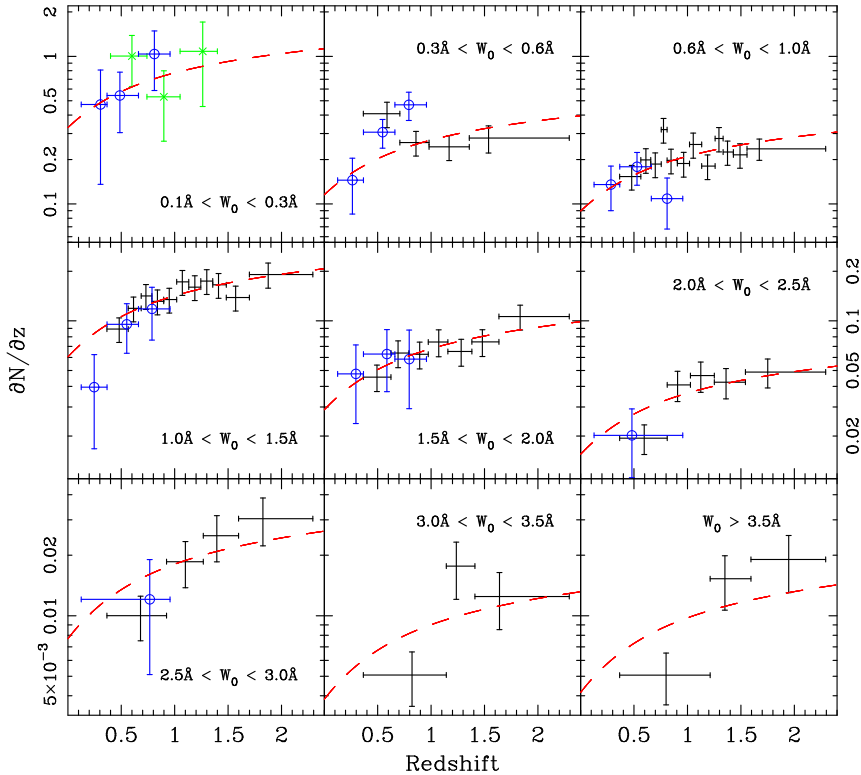}}
  \caption{Incidence as a function of $z$ for ranges of
$W_0^{\lambda2796}$.  The horizontal bars represent the redshift-bins
and the vertical bars the one-sigma uncertainty.  Circles are from the
MMT survey, and $\times$-symbols from CRCV99.  The other points
represent the EDR results.  Also shown as the dashed lines is
$\partial N/\partial z$ for a non-evolving population in a
($\Omega_{\Lambda}$, $\Omega_{M}$, $h$) = ($0.7$, $0.3$, $0.7$)
cosmology, normalized to the EDR (or CRCV99 in the top-left panel)
results.}
\label{fig:dndz}
\end{figure}

The same results on incidence presented above can be shown as a function
of redshift for $W_0^{\lambda2796}$ ranges as opposed to as a function
of $W_0^{\lambda2796}$ for a range of $z$.  This is shown in
Figure~\ref{fig:dndz} for the results of the SDSS EDR, MMT and CRCV99
studies.

There are two notable trends apparent in Figure~\ref{fig:dndz}.
First, $\partial N/\partial z$ is consistent with no evolution over a
long period of cosmic time, to within strong limits, for
intermediate-strength (0.3\AA\ $\lesssim W_0^{\lambda2796} \lesssim
2.0$) systems.  The redshift interval $[2:0.4]$, for example,
corresponds to 6 Gyrs.  There are various possible interpretations of
this result: (a) the timescales/lifetimes governing the structures are
$\gtrsim6$ Gyrs; (b) the creation/destruction of the structures is
regulated by feedback, producing a nearly steady-state; or (c) a range
of formation epochs, together with evolution, conspire to keep the
total absorption cross-section constant.  Several authors (e.g., SS92;
\cite{MM96}; CV01) have discussed the difficulties associated with
long lifetimes from effects including evaporation, hydrodynamic and
Rayleigh-Taylor instabilities, and cloud-cloud collisions.  Thus,
feedback and creation/destruction balance have been favored over
long-lifetimes (see, e.g., NTR05a).  Balanced formation/evolution
without feedback seems somewhat contrived, though it should be noted
that this is precisely what is predicted for the low-mass end of the
halo mass function in $\Lambda$CDM simulations (see, e.g.,
\cite{Reed03}).  However, without a clear picture of the physical
nature of the absorbers, it is difficult to completely eliminate any
of the scenarios.  Finally, Figure~\ref{fig:dndz} indicates that the
total absorption cross-section for systems with
$W_0^{\lambda2796}\gtrsim 2$\AA\ decreases with decreasing redshift,
especially at $z\lesssim 1$.  The evolution is such that the rate is a
function of $W_0^{\lambda2796}$, or a steepening of $\partial
N/\partial W_0^{\lambda2796}$ with decreasing $z$.  This can be
understood if: (1) various physically-distinct structures contribute
to the incidence of Mg\,II absorbers; (2) their relative contributions
are  $W_0^{\lambda2796}$-dependent; and (3) they evolve in time at
different rates.   Thus, determining the relative contribution of each
type as a function of $W_0^{\lambda2796}$ can tell us about the time
evolution of each type (and perhaps vice-versa.)

\section{Direct Study of Individual Absorbers}
Individual Mg\,II systems can be studied in more depth using
high-resolution spectroscopy to uncover the line-of-sight kinematic
structure of the absorption.  Much of these results are due to
Churchill and collaborators (\cite{CSV96}; \cite{CC98}; \cite{C00};
CV01) who have reported that: (a) intermediate-strength ($0.3 \lesssim
W_0^{\lambda2796}\lesssim 1.5$\AA) systems consist of multiple
kinematic  subsystems usually containing a dominant group and weak
kinematic outliers;  (b) such systems have multiphase ionization
structures; (c) disk+halo models have partial/limited success
describing the absorption kinematics; and (d) such systems show no
correlation between gas kinematics and galaxy properties.  Also,
direct study of galaxies associated with the absorber is possible at
relatively low ($z \lesssim 1$) redshift.  Such work in the 1990s
(e.g., \cite{BB91}; \cite{LBBD97}) found that intermediate-strength
Mg\,II absorber galaxies span a range of properties and impact
parameters consistent with normal, relatively luminous ($L \gtrsim 0.3
L^*$) galaxies.   In addition, authors (e.g., \cite{S02}) have
combined imaging and galaxy spectroscopy with high-resolution
spectroscopy of the absorption (though for only a small number of
systems that were exclusively nearly edge-on spirals), finding that
extrapolated disk rotation is consistent with the absorption
kinematics, though could not explain it in full.  Furthermore, work
presented at this symposium (e.g., Tripp 2005; Kacprzak et al. 2005)
gave new clues to the nature of Mg\,II absorbers.  For example, the
covering factor for intermediate-strength systems may be much smaller
than previously thought ($f_c\approx 0.5$ as opposed to $f_c\approx
1$) and absorption properties show no correlation with galaxy
orientation, but {\it are} correlated with galaxy asymmetry.

\subsection{A New Kinematic Regime}
\begin{figure}
\centerline{ \includegraphics[scale=0.38]{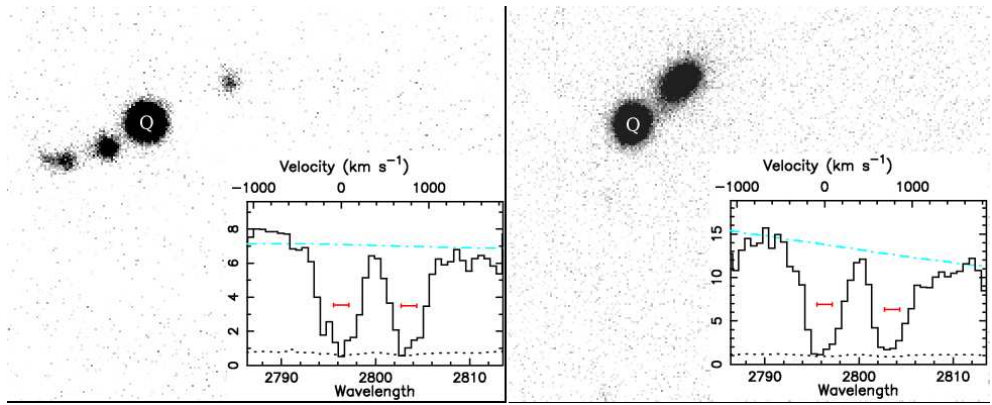}}
  \caption{``Ultra-strong'' Mg\,II absorber fields.  Quasars are
marked with a ``Q''.  Insets show the the SDSS spectrum of the Mg\,II
$\lambda\lambda2796,2803$ absorption doublet with the FWHM resolution
marked.  Left:  field of 0902$+$372, with $W_0^{\lambda2796}= 4.1$\AA\
and $z=0.670$.  If at $z=z_{abs}$, the three galaxies have $L\approx
1.2 L^*$, $b=35$ kpc (compact galaxy close to LOS), $L\approx 0.8 L^*$
and $b=72$ kpc (disturbed companion); and $L\approx 0.4 L^*$ and
$b=81$ kpc.  Right: field of 1520$+$611 with $W_0^{\lambda2796}=
4.2$\AA\ and $z=0.423$.  If at $z=z_{abs}$, the galaxy has $L\approx 2
L^*$ and $b=20$ kpc.}
\label{fig:images}
\end{figure}

Until recently, very few strong ($W_0^{\lambda2796}\gtrsim 2$\AA)
Mg\,II absorbers with relatively low-redshift were known, and no
``ultra-strong'' systems with $W_0^{\lambda2796}\gtrsim 4$\AA\ were
known.  Thus, the nature of the extreme end of the $W_0^{\lambda2796}$
distribution, which exhibits mostly saturated absorption over a {\it
huge} kinematic spread ($\approx 500$ km\,s$^{-1}$), has not been
investigated.  Our surveys with the SDSS data, however, have uncovered
a large number of such systems at redshifts low enough for direct
study.  Figure~\ref{fig:images} shows the sightlines toward two such
systems.
The 0902$+$372 field appears to contain an interacting group of three
galaxies spanning a projected distance of $\approx 150$ kpc.  There
are other galaxies at larger impact parameters, spanning $\approx500$
kpc, including a pair of interacting spirals and a large bright
galaxy. These galaxies that have magnitudes which are  consistent with
$L \approx 1-3 L^*$ if they are at the absorption redshift, suggesting
that the three galaxies  are part of a larger group.  The apparent
interaction is evidence for gas-stripping being responsible for the
extreme velocity spread seen in absorption.  The 1520$+$611 field,
contrastingly, reveals a lone bright  galaxy at lower impact
parameter.  Spectroscopy of the galaxy will be necessary to test
models based on superbubbles or rotating disk/halo absorption.

\section{Conclusions}
Evidence from both large statistical surveys and smaller
direct-observation studies provide evidence that various physically
distinct types of systems give rise to intervening Mg\,II absorbers.
Single-cloud weak systems are likely associated with enriched
Ly$\alpha$ clouds, as proposed by RCC02, in the vicinity of galaxies.
The nature of the more kinematically complex systems has been
difficult to determine with confidence.  It seems that disks and halo
clouds do contribute, though cannot explain the kinematics in full.
We know that a significant fraction of the strong Mg\,II absorbers
harbor damped Ly$\alpha$ systems; this has been discussed separately
by Rao (2005) and Turnshek et al. (2005) at this symposium.
Interaction-stripped gas and superwinds may have important
contributions, especially at large values of $W_0^{\lambda2796}$.  It
is intriguing that the various populations contribute to the incidence
of these absorbers such that $\partial N/\partial W_0$ is described
extremely well by a single exponential out to the largest
$W_0^{\lambda2796}$ values detected in recent large surveys.  Detailed
study of a large, uniform, unbiased sample of systems is necessary to
determine the relative contributions, which will hopefully lead to a
consistent understanding of their evolution in time with the
determined nature of $\partial^2 N/\partial z \partial W_0$.

\begin{acknowledgments}
We thank members of the SDSS collaboration who made the SDSS  a
success.  We acknowledge support from NSF, NASA-LTSA, and NASA-STScI
on various projects to study QSO absorption lines.  Funding for
creation and distribution of the SDSS Archive has been provided by the
Alfred P. Sloan Foundation, Participating Institutions, NASA, NSF,
DOE, the Japanese Monbukagakusho, and the Max Planck Society.
\end{acknowledgments}

\begin{multicols}{2}

\end{multicols}

\end{document}